\documentclass[aps,prl,superscriptaddress,twocolumn,showpacs,footinbib]{revtex4}

\usepackage{natbib}
\usepackage[dvips]{graphicx}
\usepackage{amsmath}
\usepackage{amsfonts}
\usepackage{subfigure}
\usepackage{psfrag}
\usepackage{color} 

\newcommand{\be}{\begin{equation}}
\newcommand{\ee}{\end{equation}}
\newcommand{\bea}{\begin{eqnarray}}
\newcommand{\eea}{\end{eqnarray}}
\newcommand{\eps}{\epsilon}
\begin{document} 

\title{Suppression of Conductance in a Topological Insulator Nanostep
  Junction}

\author{M. Alos-Palop}
\affiliation{Delft University of Technology, Kavli Institute of
  Nanoscience, Department of Quantum Nanoscience, Lorentzweg 1, 2628
  CJ Delft, The Netherlands.}
\author{Rakesh P. Tiwari}
\affiliation{Department of Physics, University of Basel,
  Klingelbergstrasse 82, CH-4056 Basel, Switzerland}
\author{M. Blaauboer}
\affiliation{Delft University of Technology, Kavli Institute of
  Nanoscience, Department of Quantum Nanoscience, Lorentzweg 1, 2628
  CJ Delft, The Netherlands.}

\date{\today}

\begin{abstract}
We investigate quantum transport via surface states in a nanostep
junction on the surface of a 3D topological insulator that involves
two different side surfaces. We calculate the conductance across the
junction within the scattering matrix formalism and find that as the
bias voltage is increased, the conductance of the nanostep junction is
suppressed by a universal factor of 1/3 compared to the conductance 
of a similar planar junction based on a single surface of a topological 
insulator. We also calculate and analyze the Fano factor of the
nanostep junction and predict that the Fano factor saturates at
$1/5$, five times smaller than for a Poisson process.
\end{abstract}

\maketitle

Experimental demonstration of topological phases in both
two-dimensional (such as HgTe) and three-dimensional (such as
Bi$_2$Se$_3$) compounds with strong spin-orbit interaction
~\cite{Konig2007,Hsieh2008,Xia2009a,Xia2009b} has generated a plethora
of interest in the physics community~\cite{Hasan2010}.  These
compounds are insulating in the bulk (since they have an energy gap
between the conduction band and the valence band) but their surfaces
support gapless topological excitations. These surface states are
topologically protected against non-magnetic defects by time-reversal
symmetry~\cite{Hasan2010}.

\begin{figure}[h]
\includegraphics[width=.8\columnwidth]{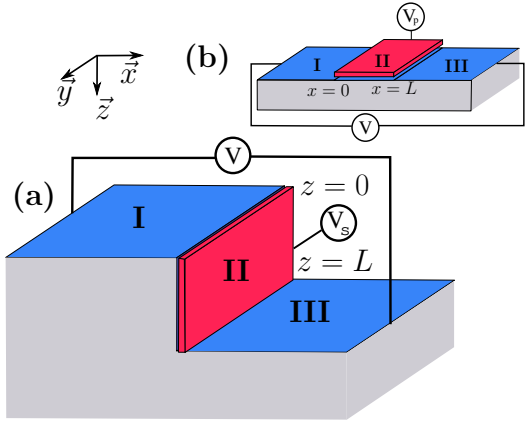}
\caption{Schematics of the proposed junction. (a) The nanostep
  junction.  As the name suggests the height $L$ of the junction is
  $\sim$ 10~nm. (b) Similar junction, involving only one side surface
  of the topological insulator. See the text for further details.}
\label{fig:sketchStep}
\end{figure}

In the simplest case these low-energy excitations of a strong
three-dimensional (3D) topological insulator can be described by a
single Dirac cone at the center of the two-dimensional Brillouin zone
($\Gamma$ point) ~\cite{Konig2007,Xia2009a,Hasan2010}. Recently,
naturally occurring defects such as step-like
imperfections~\cite{Seo2010, Alpichshev2010} have been studied in the
context of topological insulators. Conductance through atomic scale
step-defects has been measured demonstrating that the surface states
are protected in spite of such an abrupt defect~\cite{Seo2010}. These
\textit{atomic} sized steps are usually modeled as delta function
potential barriers~\cite{StepDefects}. In this article we investigate
quantum transport through a \textit{nanostep} junction involving two
different side surfaces of a 3D topological insulator. We predict that
the conductance of the nanostep junction is suppressed in the
large-energy limit by a universal factor of 1/3 as compared to the
conductance of a similar junction based on a single surface of a 3D
topological insulator (or a similar junction in
graphene). Figure~\ref{fig:sketchStep}(a) shows a schematic of the
nanostep junction considered. The junction is divided into three
regions. In region I ($x < 0$ and $z=0$) and in region III ($x>0$
and $z=L$) the surface states lie in the $x$-$y$ plane, while in
region II ($x=0$ and $0 \leq z \leq L$) the surface states lie in the
$y$-$z$ plane. A dc bias voltage is applied between region I and III
and a top gate $V_s$ controls the carriers in region
II. Figure~\ref{fig:sketchStep}(b) shows a schematic of the analogous
planar junction, where in all three regions the surface states lie
in the $x$-$y$ plane and a top gate $V_p$ is applied to region II. The
first junction [Fig.~\ref{fig:sketchStep}(a)] is referred to as a {\it
  step junction}, and the second junction
[Fig.~\ref{fig:sketchStep}(b)] as a {\it planar junction} in the rest
of this article.  We use the conceptually transparent scattering
matrix formalism to calculate the transport properties of these
junctions~\cite{yuli}. For concreteness, we use typical parameter
values of the 3D topological insulator Bi$_2$Se$_3$ when illustrating
our results.

The low-energy effective Hamiltonian for Bi$_2$Se$_3$ in the basis of
four hybridized states of Se and Bi $p_z$-orbitals denoted as
$\{\left\vert p1_{z}^{+},\uparrow \right\rangle $, $\left\vert
p2_{z}^{-},\uparrow \right\rangle $, $\left\vert p1_{z}^{+},\downarrow
\right\rangle $, $\left\vert p2_{z}^{-},\downarrow \right\rangle \}$
can be written as~\cite{wen2010}:
\begin{equation}
H(\mathbf{k})=\epsilon _{0}(\mathbf{k})+\left[
\begin{array}{cccc}
\mathcal{M}(\mathbf{k}) & A_{1} k_z & 0 & A_{2}k_{-} \\
A_{1} k_z  & -\mathcal{M}(\mathbf{k}) & A_{2}k_{-} & 0 \\
0 & A_{2}k_{+} & \mathcal{M}(\mathbf{k}) & -A_{1} k_z \\
A_{2}k_{+} & 0 & -A_{1} k_z & -\mathcal{M}(\mathbf{k})%
\end{array}
\right] ,
\label{eq:HZhang}
\end{equation}
where $k_{\pm } = k_{x} \pm i k_{y}$, $\epsilon_{0} (\mathbf{k}) = C +
D_{1} k_{z}^{2} + D_{2} k_{+} k_{-}$, $\mathcal{M} (\mathbf{k}) = M -
B_{1} k_{z}^{2} - B_{2} k_{+} k_{-}$, and $k_{+} k_{-} =k_{x}^{2} +
k_{y}^{2}$~\cite{parameters}. Here $\uparrow$ $(\downarrow)$ stands
for up (down) spin and $+$ $(-)$ stands for even (odd) parity.  From
this three-dimensional Hamiltonian, there exists a straightforward
procedure to obtain the effective Hamiltonian describing the surface
states \cite{Qui2011,wen2010}. The surface states in the $x$-$y$
plane, for example, are obtained from the three-dimensional
wavefunctions for these surface states (which are exponentially
damped in the $z$-direction, with finite skin depth $\lambda$) using
Eq.~(\ref{eq:HZhang}), followed by imposing the boundary conditions of
vanishing wavefunctions at the two boundaries ($z=0$ and $z=L$). For
three-dimensional topological insulators ($L\gg\lambda$) the surface
states at the two boundaries are decoupled and the effective
Hamiltonian describing the carriers in regions I and III is then given
by~\cite{Qui2011}:
\begin{equation}
 \mathcal{H}^{xy} = \epsilon_0^{xy} + \hbar v_{F}^{xy}
 (\sigma_xk_y-\sigma_yk_x),
\label{eq:hxy}
\end{equation}
 where $\epsilon_0^{xy}=C+\frac{D_1}{B_1}M$, $\hbar
 v_{F}^{xy}=A_2\sqrt{1-\frac{D_1^2}{B_1^2}}$ represents the Fermi
 velocity in the $x$-$y$ plane and $\sigma_x, \sigma_y$ and $\sigma_z$
 denote the usual Pauli matrices.

Analogously, we obtain the effective Hamiltonian describing the
carriers in region II as
\begin{equation}
  \mathcal{H}^{yz} = \epsilon_0^{yz} + eV_s + \hbar v_{F}^{yz}
  (\sigma_y \frac{A_1}{A_2}k_z-\sigma_zk_y),
\label{eq:hyz}
\end{equation}
where $\epsilon_0^{yz}=C+\frac{D_2}{B_2}M$ and $\hbar
v_{F}^{yz}=A_2\sqrt{1-\frac{D_2^2}{B_2^2}}$. Solving the 
Hamiltonian~(\ref{eq:hxy}), we obtain the eigenstates in
region I and region III as
\begin{eqnarray}
\Psi^{\pm}_{\text{I}(\text{III})
} &=& \frac{1}{\sqrt{2}}\left[
\begin{array}{c}
1  \\
\mp i e^{\pm i \phi}
\end{array}%
\right] e^{ i k_y y}e^{\pm i k_x x },
\end{eqnarray}
with corresponding energy eigenvalues given by $\epsilon =
\epsilon_0^{xy} + \hbar v_{F}^{xy} \sqrt{k_x^2 + k_y^2}$ and
$\tan(\phi) \equiv k_y / k_x$. Similarly, in region II we obtain:
\begin{eqnarray}
\Psi^{\pm}_{\text{II}} = \frac{1}{\sqrt{2 (1 + \sin(\gamma))}} \left[
\begin{array}{c}
\mp  i \cos(\gamma) \\
1+\sin(\gamma)
\end{array}%
\right] e^{i k_y y }e^{\pm i k_z z},
\end{eqnarray}
with corresponding energy eigenvalues $\epsilon~=~\epsilon_0^{yz} + 
eV_s + \hbar v_{F}^{yz} \sqrt{k_y^2 +
  (A_1^2/A_2^2) k_z^2}$ and $\tan(\gamma) \equiv A_2 k_y / (A_1 k_z)$.
The $+(-)$ labels of the wavefunction indicate right (left) traveling
carriers in regions I and III, and downwards (upwards) traveling
carriers in region II. It should be noted that in general
$\epsilon_0^{xy}\neq \epsilon_0^{yz}$, which implies that the Dirac
cone describing the surface states in the $y$-$z$ plane is shifted by
an energy of $\epsilon_0 \equiv \epsilon_0^{yz}-\epsilon_0^{xy}$ with
respect to the Dirac cone describing surface states in the $x$-$y$
plane. Furthermore, the Dirac cone describing the excitations in
region II has elliptic cross section ($A_1\neq A_2$). These features
are in good agreement with recent electronic structure calculations of
similar systems~\cite{Moon2011}.

Considering electrons incident from left to right, the total
wavefunction in the different regions can be written as:
\begin{figure}
\subfigure{\includegraphics[width=0.7\columnwidth]{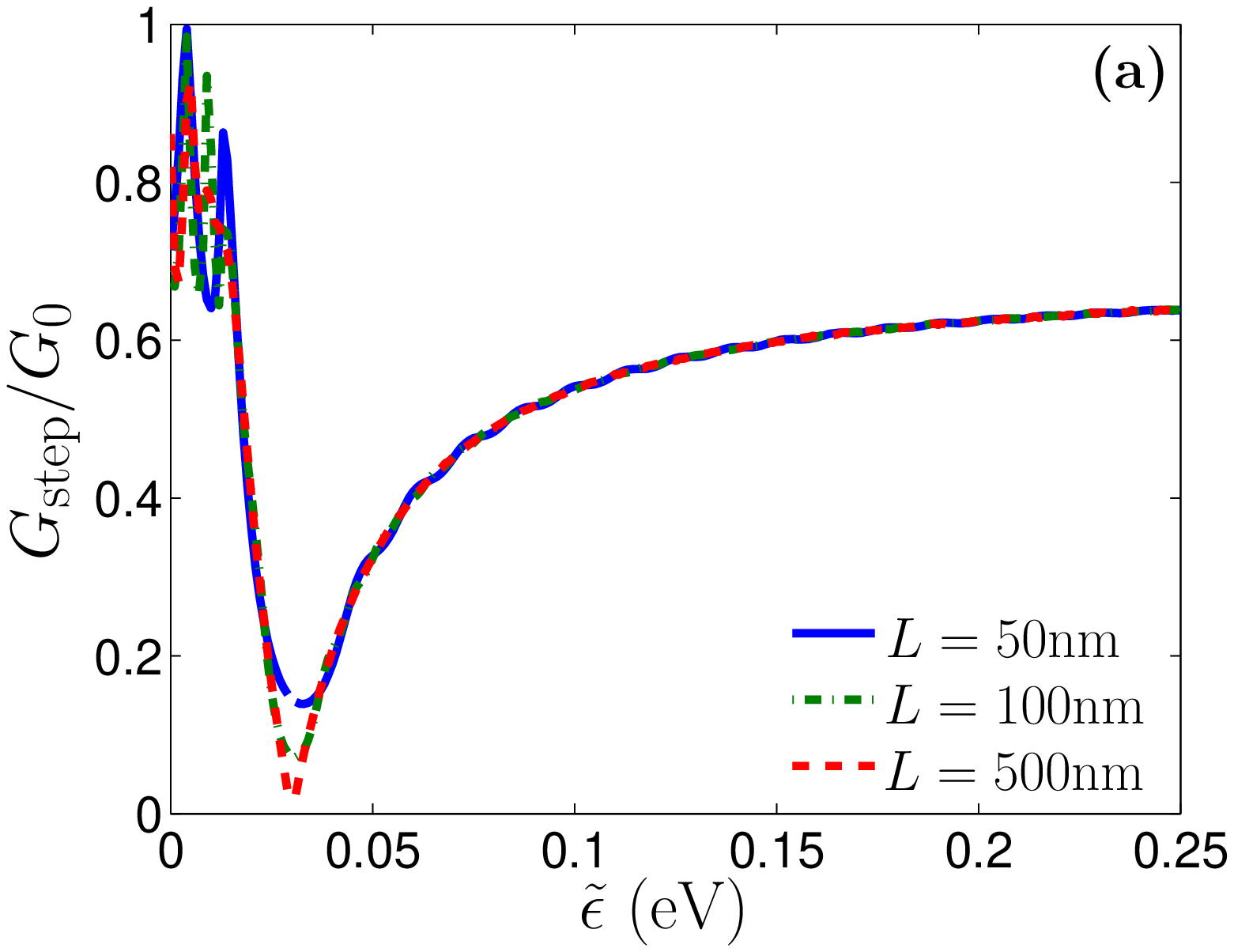}}

\subfigure{\includegraphics[width=0.7\columnwidth]{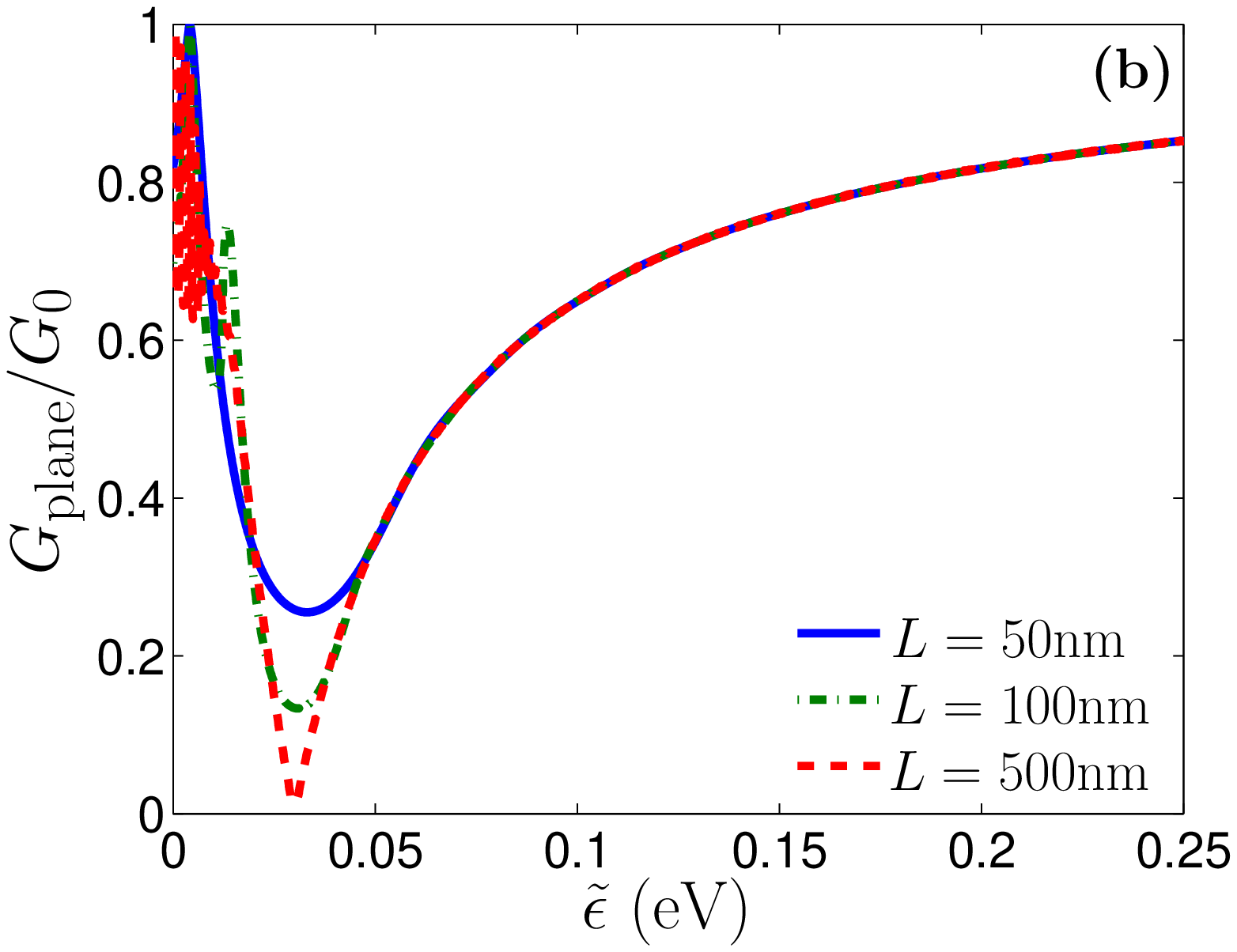}}
\caption{(Color online) Conductance as a function of energy for (a)
  the step junction [Eq.~(\ref{eq:cond_step})] and (b) the planar junction
  [Eq.~(\ref{eq:PlaneT})]. Different junction widths are plotted. In
  both plots, the solid (blue) line denotes $L=50$ nm, the dot-dashed
  (green) line denotes $L=100$ nm, and the dashed (red) line
  corresponds to $L=500$ nm. Parameters used are $A_1=2.2$~eV$\AA$,
  $A_2=4.1$~eV$\AA$, $\hbar v_F^{xy} =4.065$~eV$\AA$ , $\hbar v_F^{yz}
  =3.846$~eV$\AA$, $\eps_0^{xy}=0.03$~eV, $\eps_0^{yz}=0.09$~eV,
  $eV_s=-0.03$~eV and $eV_p=0.03$~eV. See the text for more details.}
\label{fig:conductance}
\end{figure}
\begin{equation}
\begin{cases}
\Psi_{\text{I}} = \Psi^{+}_{\text{I}} + r \Psi^{-}_{\text{I}} &
\text{if } x \leq 0, \: z=0,\\
\Psi_{\text{II}} = a \Psi^{+}_{\text{II}} + b \Psi^{-}_{\text{II}} &
\text{if } 0\leq z \leq L, \: x=0,\\
\Psi_{\text{III}} = t \Psi^{+}_{\text{III}} & \text{if } x \geq 0, \: z=L .
\end{cases}
\label{eq:TotalWaveF}
\end{equation}
The reflection and transmission coefficients $r$ and $t$ can be obtained
by imposing the boundary conditions under which the current normal to
the boundary is conserved~\cite{BoundaryGraphene,Sen2012,boundary}.
We then find for the transmission
probability $T \equiv t^{\ast}t$ of an electron incident on the step
junction at a given angle of incidence $\phi$:
\begin{equation}
T_{\text{step}} (\phi)= \frac{\cos^2(\phi) \cos^2(\gamma)}{\cos^2(\phi) 
\cos^2(\gamma) \cos^2 (k_z L) + \sin^2 (k_z L)}
\label{eq:StepT}
\end{equation}
with 
\be \sin(\gamma) = \kappa \sin(\phi),
\label{eq:singamma}
\ee 
$\kappa = \left( \frac{v_F^{yz}}{ v_F^{xy} } \right) \left(
\frac{\tilde{\epsilon} }{\tilde{\epsilon} - \eps_0 - eV_s}\right) $
and $\tilde{\epsilon} \equiv \epsilon -
\epsilon^{xy}_{0}$. Eq.~(\ref{eq:singamma}) is obtained by using
conservation of energy and conservation of momentum along the
$y$-direction.

The zero-temperature conductance $G_{\text{step}}$ across the step
junction is then given by~\cite{yuli}:
\begin{equation}
G_{\text{step}} = G_0\, \int^{\pi/2}_0 d\phi\, T_{\text{step}}(\phi)
\cos(\phi).
\label{eq:cond_step}
\end{equation}
Here $G_0 \equiv \frac{2e^2}{h}\rho(\tilde{\epsilon})\hbar v_F W$,
$\rho(\tilde{\epsilon})=\tilde{\epsilon} /(\pi(\hbar v_F)^2)$ denotes
the density of states, $W$ the sample width, and the integration is
over all angles of incidence $\phi$. Figure~\ref{fig:conductance}(a)
shows the conductance $G_{\text{step}}$ [Eq.~(\ref{eq:cond_step})] as
a function of energy $\tilde{\epsilon}$ for different values of
$L$. We observe that the conductance reaches a minimum at
$\tilde{\epsilon} = \epsilon_{0}+ eV_s$, arising from the difference
in Dirac point energies in different planes. Beyond this point, the
conductance first oscillates and then saturates at large energies. For
$\kappa~\rightarrow~1$ the conductance $G_{\text{step}}$ reaches a
limiting value of 2/3, independent of barrier width $L$. Below (see
Eq.~(\ref{eq:translimit})) we discuss the behavior of the conductance
close to this saturation point in more detail.

The transmission probability of the analogous planar junction [see 
Fig.~\ref{fig:sketchStep}(b)] with a top 
gate in the middle region is given by~\cite{vBarrierGraphene}:
\begin{equation}
T_{\text{plane}}(\phi) = \frac{1}{ \cos^2 (k'_x L) + \sin^2 (k'_x L)
  \frac{(1 - \sin(\phi )\sin (\gamma^{\prime}))^2}{\cos^2(\phi)
    \cos^2(\gamma^{\prime})} }.
\label{eq:PlaneT}
\end{equation}
Here $k'_x$ represents the $x$-component of the momentum in region II,
the energy dispersion is given by $\epsilon~=~\epsilon_0^{xy} +eV_p+
\hbar v_{F}^{xy} \sqrt{k_x^{'2} + k_y^2}$ and $\gamma^{\prime} \equiv
\tan^{-1}(\frac{k_y}{k'_x})$.
Figure~\ref{fig:conductance}(b) shows the conductance
$G_{\text{plane}}$ as a function of energy $\epsilon$ for different
values of $L$.  As before, the conductance reaches a minimum when
$\tilde{\epsilon} =eV_p$, and then increases to reach its saturation value
$G/G_0 = 1$. For larger energies, the conductance of the step junction
is thus suppressed by a factor of 1/3 compared with planar
junctions. We attribute this suppression to the fact that the carriers
in the step junction \textit{have to} change their plane of
propagation in region II~\cite{spinorbit}.  

\begin{figure}
\subfigure{\includegraphics[width=.49\columnwidth]{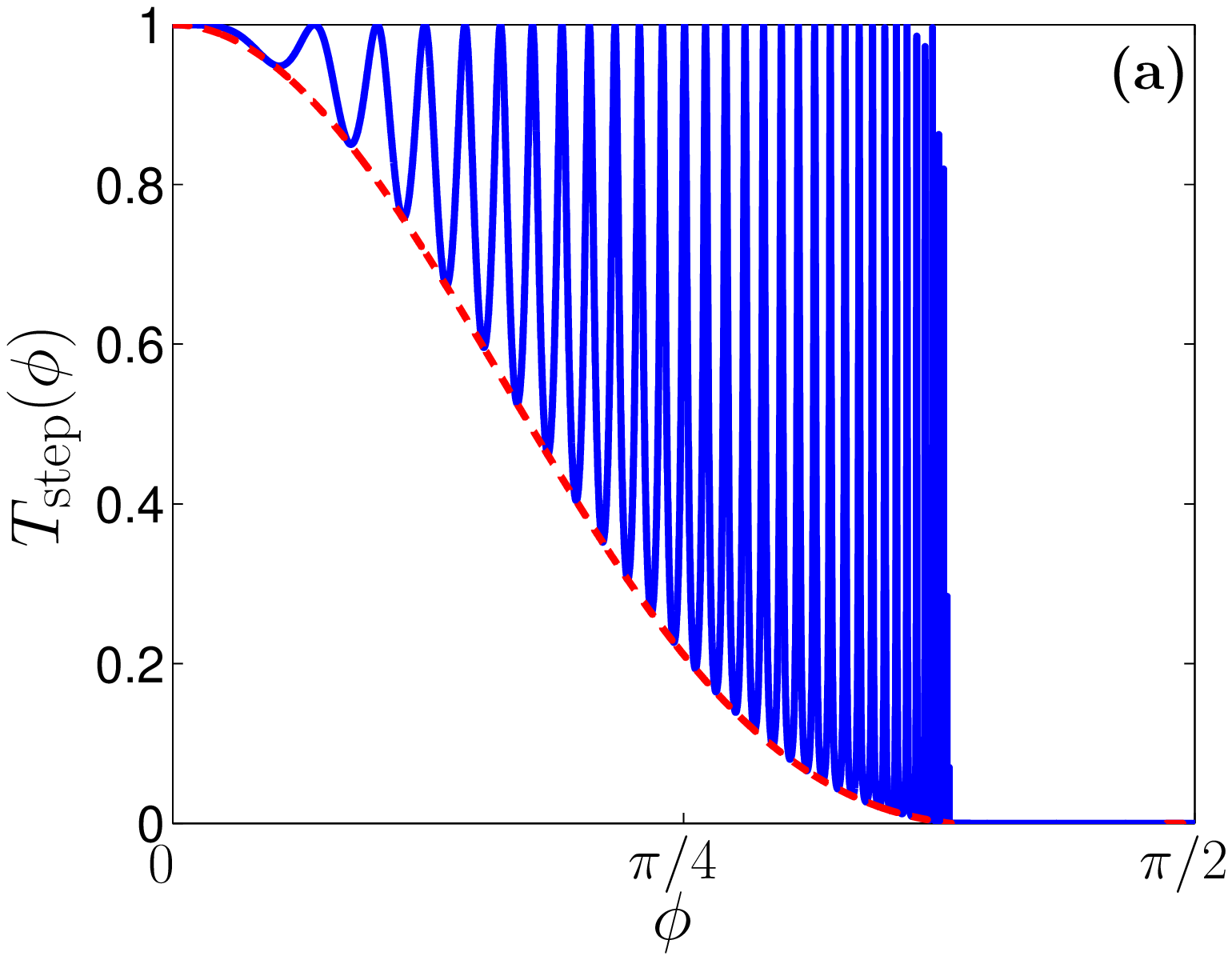}}
\subfigure{\includegraphics[width=.49\columnwidth]{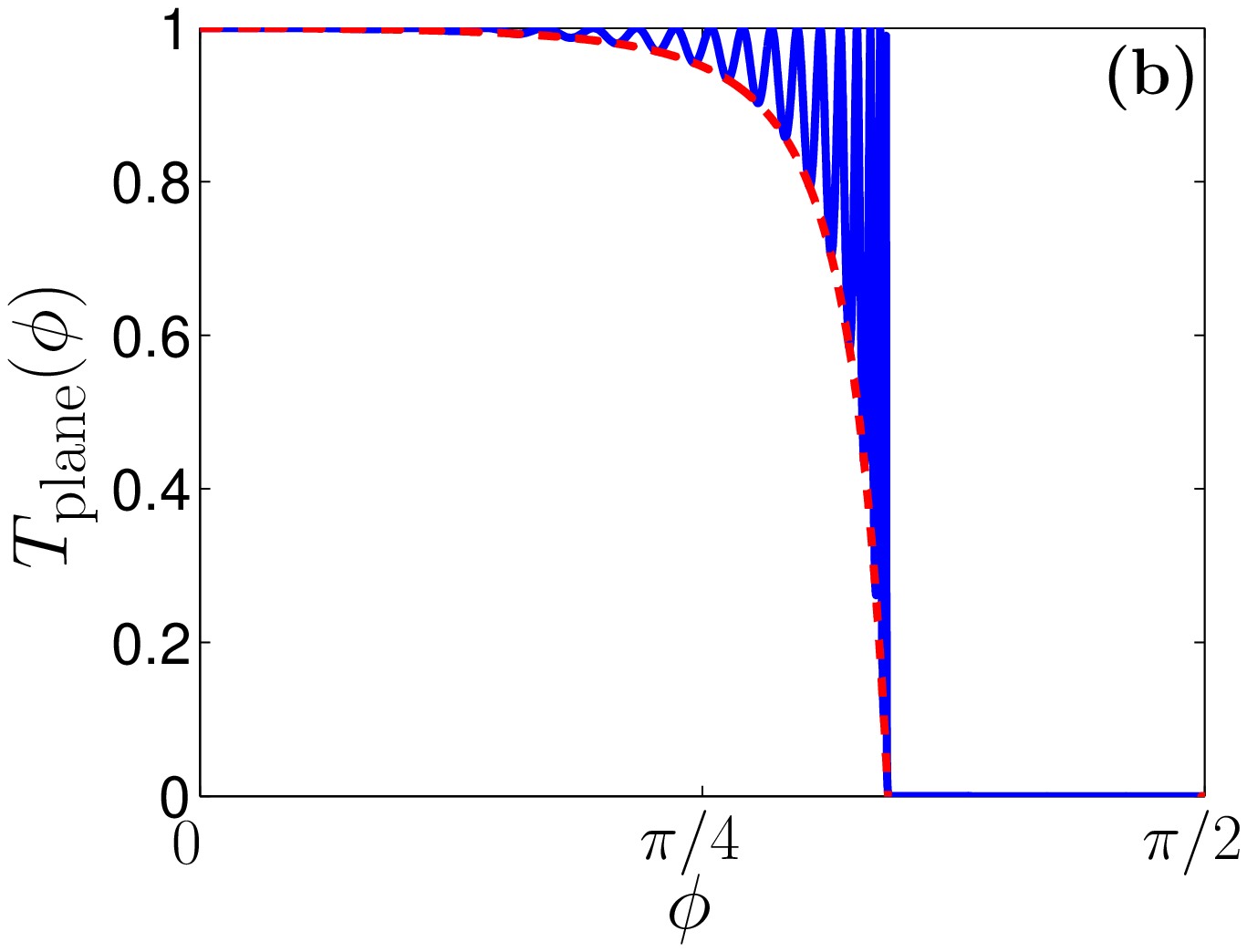}}
\caption{(Color online) Transmission (a) $T_{\text{step}}(\phi)$
  [Eq.~(\ref{eq:StepT})] for the step junction and (b)
  $T_{\text{plane}}(\phi)$ [Eq.~(\ref{eq:PlaneT})] for the planar
  junction as a function of angle of incidence $\phi$ for $L=100$~nm
  and $\tilde{\epsilon}=0.25$~eV. Parameters used are the same as in
  Fig.~\ref{fig:conductance}. The dashed red lines mark
  the envelopes of the transmission probabilities.}
\label{fig:transmission}
\end{figure}
We now analyze in more detail the difference between the conductance
of the step and the planar junction by comparing the denominators in
Eqns.~(\ref{eq:StepT})
and~(\ref{eq:PlaneT}). Figure~\ref{fig:transmission} shows the
transmission probabilities $T_{\text{step}}(\phi)$ and
$T_{\text{plane}}(\phi)$ as a function of the angle of incidence
$\phi$. From Fig.~\ref{fig:transmission}(a) we see that for the step
junction there is a cut-off angle of incidence, which arises from the
finite energy and velocity mismatch. This \textit{critical} angle can
be expressed as $\phi_{\text c, step} = \sin^{-1} \left(
\frac{v_{F}^{xy}} {v_{F}^{yz}} \frac{\tilde{\epsilon} - \epsilon_0
  -eV_s}{\tilde{\epsilon}} \right)$.  We also see that the conductance
of both junctions includes contributions from \textit{many} resonant
modes. Here, a resonant mode is defined as a mode with an angle of
incidence for which the transmission $T=1$. The various minima of
these resonant modes form an envelope, as shown by the dashed (red)
lines in Fig.~\ref{fig:transmission}. These envelope functions are
obtained from the transmission expressions Eqns.~(\ref{eq:StepT}) and
(\ref{eq:PlaneT}) by setting $k_zL=(2n+1)\frac{\pi}{2}$ and
$k'_xL=(2n+1)\frac{\pi}{2}$, respectively, with $n$ integer. When the
number of resonant modes is large ($n\gg1$) a lower bound for
the integrated transmission is obtained by integrating over the
envelope function. Under these conditions the
conductance $G_{\text{step}}$ of the step junction
[Eq.~(\ref{eq:cond_step})] becomes
\begin{eqnarray}
G_{\text{step}}/G_0 & \stackrel{n\gg1}{\rightarrow} & \int^{\pi/2}_0
d\phi\, \cos^3(\phi) \cos^2(\gamma) \nonumber \\ 
& = & \frac{2}{3}-\frac{2}{15}\kappa^2.
\label{eq:translimit}
\end{eqnarray}
Similarly, we find for the planar junction
\begin{equation}
G_{\text{plane}}/G_0 \stackrel{n\gg1}{\rightarrow}
\frac{\sqrt{\delta}(3\delta -1)+(\delta-1)^2
  \tanh^{-1}(\sqrt{\delta})} {2\, \delta^{3/2}},
\end{equation}
where $\delta \equiv
\frac{\tilde{\epsilon}}{\tilde{\epsilon}-eV_{p}}$. The suppression of
the conductance $G_{\text{step}}$ in the vicinity of the saturation
value thus depends on the shift of the Dirac point energies and the
ratio of the Fermi velocities in region I and II~\cite{exactbound}.
As final remarks, we note that the effect of the elliptical dispersion
in the middle region of the step junction can be incorporated as an
effectively wider barrier, $\tilde{L}=L \frac{A_2}{A_1}$ and
$\frac{A_2}{A_1}>1$, compared to the planar junction. The minimum
barrier width needed to observe the suppression in the conductance
described above is given by the condition for the existence of the
first resonant mode in the junction, $L =\pi/k_z$. In the case of
Bi$_2$Se$_3$, this minimum width is $L \sim 5$~nm. For cleaved
topological insulators, widths of $L \sim 1$~nm have been
reported~\cite{GZhang2009}.

\begin{figure}
\subfigure{\includegraphics[width=.7\columnwidth]{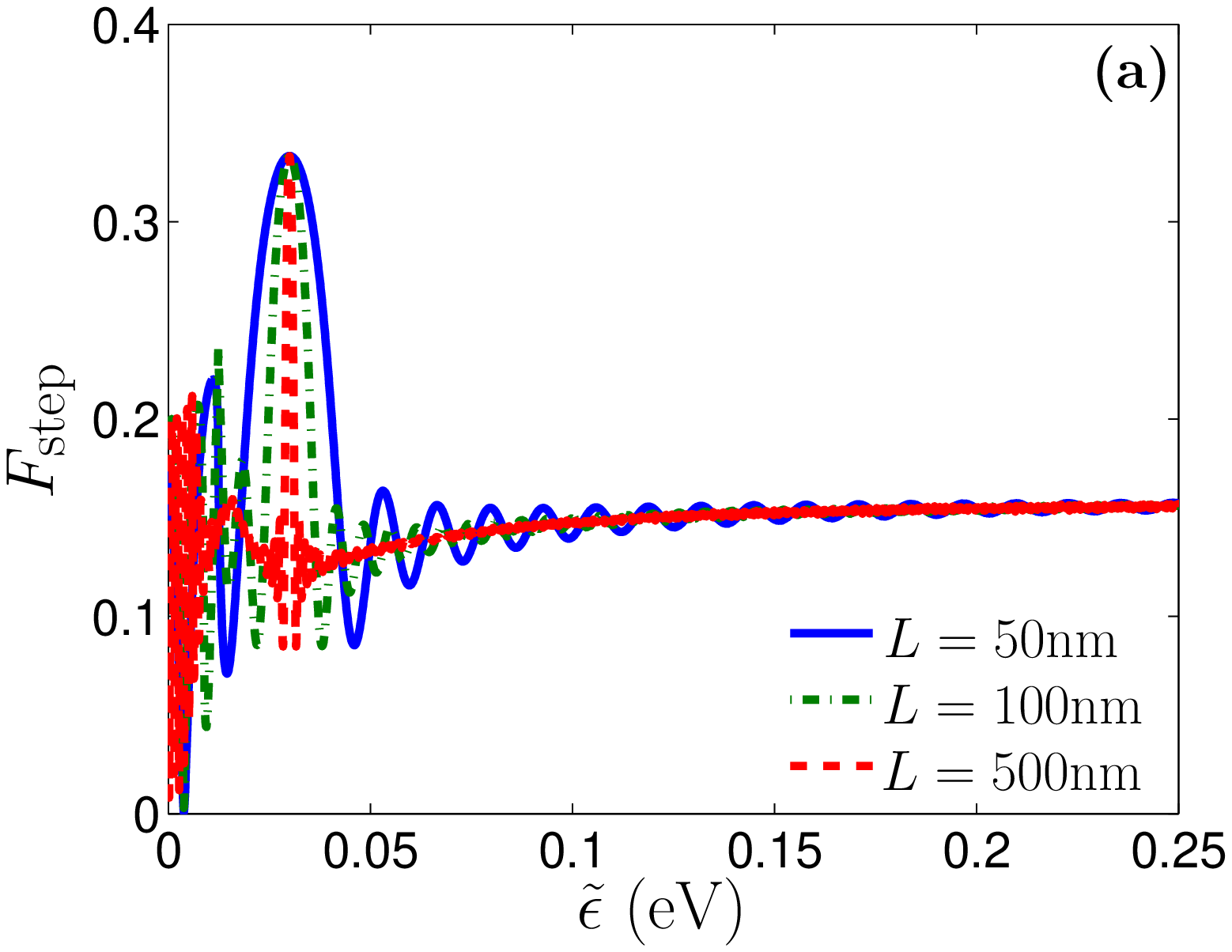}}

\subfigure{\includegraphics[width=.7\columnwidth]{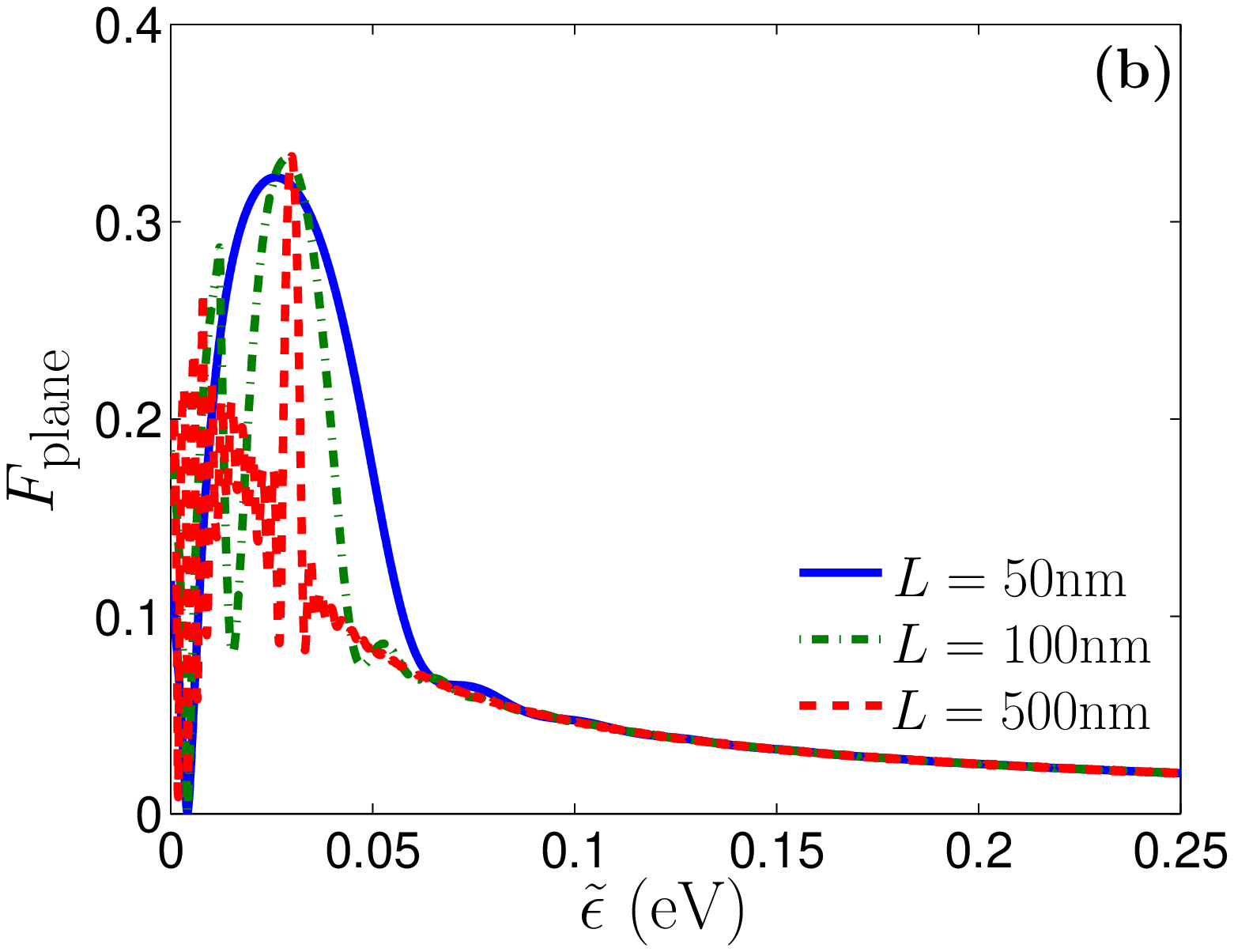}}
\caption{(Color online) The Fano factor [Eq.~(\ref{eq:Fano})] for (a)
  the step junction and (b) the planar junction as a function of
  energy $\epsilon$. Different junction widths are plotted. In both
  plots, the solid (blue) line denotes L = 50 nm, the dot-dashed
  (green) line denotes L = 100 nm, and the dashed (red) line
  corresponds to L = 500 nm.  Parameters used are the same as in
  Fig.~\ref{fig:conductance}.}
\label{fig:Fano}
\end{figure}
In the remaining part of this paper we investigate the Fano factor of
the step junction, which is a measure for the noise suppression in the
junction relative to Poisson noise~\cite{yuli}. Within the scattering
matrix formalism the Fano factor $F$ is defined as
\begin{equation}
F = \frac{\int_0^{\pi/2} d\phi~ \cos(\phi)~T (1-T)}{\int_0^{\pi/2}
  d\phi~ \cos(\phi)~T},
\label{eq:Fano}
\end{equation}
where $\phi$ is the angle of incidence of the carriers and $T$
represents the transmission of the junction. Substituting
Eqns.~(\ref{eq:StepT}) and (\ref{eq:PlaneT}) into Eq.~(\ref{eq:Fano})
and evaluating the integrals in general leads to lengthy
expressions. However, in the limit of a large number of resonant modes
($n\gg1$) we find that the Fano factor of the step junction is given
by~\cite{approx}
\begin{equation}
 F_{\text{step}} \stackrel{n\gg1}{\rightarrow} \frac{1}{5} +
 \frac{12}{175} \kappa^2,
\label{eq:Fanostep}
\end{equation}
where $\kappa$ as defined earlier.

Figure~\ref{fig:Fano} shows the calculated F$_{\text{step}}$ and
F$_{\text{plane}}$ for different values of the junction width L. As
the energy of the incident carriers increases, both Fano factors reach
a maximum around $\tilde{\epsilon} = 0.03$ eV, where $\tilde{\epsilon}
= \epsilon_0 + eV_s$ and $\tilde{\epsilon} = eV_p$ respectively. For
higher energies the Fano factor of the step junction oscillates around
its saturation value $1/5$, which is five times smaller than the Fano
factor expected for a Poisson process, $F=1$. On the other hand, the
Fano factor F$_{\text{plane}}$ for the planar junction vanishes as the
energy of the incident carriers increases. This can be explained by
noticing that the transmission T$_{\text{plane}} \rightarrow 1$ as
$\tilde{\epsilon}$ increases (see Fig.~\ref{fig:conductance}(b)). This
remarkable difference in the Fano factor for the two junctions again
suggests that there exists an additional scattering mechanism for the
step junction that does not exist in the planar junction, namely the
change of plane of propagation of the topological surface states.

In conclusion, we have proposed and analyzed quantum transport through
a nanostep junction on the surface of a 3D topological insulator in
the ballistic limit. Our results show that the conductance in a
nanostep junction is suppressed by up to a factor of 1/3 compared to
similar junctions based on a single surface of a 3D topological
insulator or graphene. Although the suppression depends on the ratio
of the Fermi velocities and the difference in Dirac point energies of
the different side surfaces of the 3D topological insulator, the
saturation values of the conductance and Fano factor themselves
G$_{\text{step}} \rightarrow 2/3$ and F$_{\text{step}} \rightarrow
1/5$ are universal. We also predict oscillating behavior of the Fano
factor around its saturation value. Experimental
demonstration~\cite{DiCarlo2008} of our predictions will provide
further insight into the scattering mechanisms involved in topological
insulator nanostep junctions.

We would like to thank C. Bruder for valuable discussions. This
research was supported by the Dutch Science Foundation NWO/FOM. RPT
acknowledges financial support by the Swiss SNF, the NCCR Nanoscience,
and the NCCR Quantum Science and Technology.


\begin{thebibliography}{}
\bibitem{boundary} Notice that in our case the current normal to the
  junction has $J_x$ and $J_z$-components and boundary conditions are
  given by $J_x\mid_{x=0^{-}}^{z=0} = J_z\mid_{x=0}^{z=0^{+}}$ and $
  J_z\mid_{x=0}^{z=L^{-}} = J_x\mid_{x=0^{+}}^{z=L}$.
\bibitem{spinorbit} For topological insulators with strong spin orbit
  interactions, a change in the plane of propagation is accompanied by
  a change in the spin due to the helical nature of the surface
  states. Although our model does not take spin explicitly into
  account, it is implicitly included in the Hamiltonian
  (\ref{eq:HZhang}).
\bibitem{exactbound}
Note that the lower bound Eq.~(\ref{eq:translimit}) becomes exact
in the limit $\kappa \rightarrow 0$.
\bibitem{approx} Note that the approximation $n \gg 1$ gives a lower
  bound on the conductance and an upper bound on the Fano factor.
\end{thebibliography}
\end{document}